# Coupling Spin Defects in Hexagonal Boron Nitride to Titanium Oxide Ring Resonators

Milad Nonahal[1], Chi Li[1,2], Febiana Tjiptoharsono[3], Lu Ding[3], Connor Stewart[1], John Scott[1,2], Milos Toth[1,2], Son Tung Ha[3], Mehran Kianinia[1,2], Igor Aharonovich[1,2*]

1. School of Mathematical and Physical Sciences, Faculty of Science, University of Technology Sydney, Ultimo, New South Wales 2007, Australia

2. ARC Centre of Excellence for Transformative Meta-Optical Systems (TMOS), University of Technology Sydney, Ultimo, New South Wales 2007, Australia

3. Institute of Materials Research and Engineering, A*STAR (Agency for Science, Technology and Research), Kinesis, 138635 Singapore

**Abstract:**

*Spin-dependent optical transitions are attractive for a plethora of applications in quantum technologies. Here we report on utilization of high quality ring resonators fabricated from TiO$_2$ to enhance the emission from negatively charged boron vacancies ($V_B^-$) in hexagonal Boron Nitride. We show that the emission from these defects can efficiently couple into the whispering gallery modes of the ring resonators. Optically coupled $V_B^-$ showed photoluminescence contrast in optically detected magnetic resonance signals from the hybrid coupled devices. Our results demonstrate a practical method for integration of spin defects in 2D materials with dielectric resonators which is a promising platform for quantum technologies.*

Defects in hexagonal boron nitride (hBN) provide a test bed for the study of light-matter interactions and nanophotonics with two-dimensional materials at room temperature [1-5]. Among them, optically addressable spin defects have recently gained momentum due to their relevant application in quantum sensing and quantum information technologies [6-10]. Particularly, negatively charged boron vacancies ($V_B^-$) is a spin 1 system which can be mapped out through optically detected magnetic resonance (ODMR) spectroscopy at room temperature [11-15]. At zero field, the transitions between two spin states ($m_s = 0$ and $m_s = \pm 1$) in the ground state result in two resonances at ~3.4 and ~3.5 GHz in the ODMR signal [11]. Spin coherence of $V_B^-$ has been exploited for high temporal and spatial resolution quantum sensing of temperature, strain, electric and magnetic fields[16-20]. However, the photoluminescence emission from $V_B^-$ spans in the NIR has no clear indication of zero phonon line (ZPL) even at cryogenic temperature, and the exact electronic level structure and emission dipole of $V_B^-$ are yet to be understood. Just recently, coupling of the defects into high-quality cavities suggests the ZPL spectral location to be around 770 nm [21] and excited state spectroscopy of the defects revealed the spin states in the excited state[22-24]. Moreover, the defect suffers from low intrinsic brightness and quantum efficiency. This limitation can be overcome by coupling the $V_B^-$ to micro-resonators that enables efficient control over the emission properties as such presents an attractive platform for enhanced understanding and utilization of spin defects in van der Waals materials. Recent attempts have been made to couple $V_B^-$ into bullseye[25] or plasmonic[26] structures which results in the enhancement of the emission. Dielectric resonators such as ring resonators supporting whispering gallery modes (WGMs), provide high-quality factors in low mode volume structures and hence are attractive alternatives [27-30].

Here we realise a hybrid approach based on hBN flake as $V_B^-$-hosting material on top of $TiO_2$ as a cavity platform. Since a high-quality cavity requires relatively large refractive index of ~2.5 and low optical loss in the wide wavelength range in the visible and IR, $TiO_2$ is an attractive material for the fabrication of waveguides and photonic resonators [31-34]. The resonator is designed to support multimode WGMs with small free spectral range (FSR) in an attempt to enhance the broad emission of $V_B^-$. Finally, the ODMR collected from the ring confirms the efficient coupling of spin defects in hBN into WGM modes of the $TiO_2$ ring resonator.

We developed a fabrication process to realize hBN/TiO2 hybrid resonator and to demonstrate emission enhancement of spin defect by coupling to the WGMs. Given the layered nature, hBN could be readily integrated with other systems by the stacking strategy[35]. However, transferring hBN on cavity structures inevitably damages these fragile structures. Thus, to avoid this, a heterostructure was prepared prior to the fabrication process. The fabrication steps are described in detail in the Method section and schematically demonstrated in figure 1a. Briefly, a 200-nm of $TiO_2$ was first sputtered onto the Si substrate through ion-assisted deposition (IAD). The resulting film showed a smooth surface (average roughness of 0.9 nm) and high refractive index (2.3 at 800 nm). The characterization of the deposited $TiO_2$ film is provided in the supporting information (figure SI-1). hBN flakes were mechanically transferred from high quality bulk crystal onto the $TiO_2$ surface via scotch tape method. The sample was then annealed at 300 °C to remove the tape residual and maximize surface adhesion to form a stable heterostructure. At this temperature, $TiO_2$ phase transition and the associated change in the refractive index can be avoided to retain the low optical loss of the $TiO_2$ [36]. To increase accessibility of the defect to the cavity in hybrid approach, a thin flake (~ 20 nm) was picked for the fabrication. Therefore, the defect could be spatially closer to the cavity resulting in improved coupling efficiency since it minimizes optical loss in the hBN material. An array of wheel-like structures was subsequently patterned on the heterostructure using EBL technique followed by deposition of 50 nm Cr. After liftoff, the remaining Cr acting as a metal mask for the etching step. The pattern was then transferred into the heterostructure by reactive ion etching (RIE) method (figure 1a-i). Following the RIE step, potassium hydroxide (KOH) aqueous solution was used to selectively etch the underlying Si substrate as shown in figure 1a-ii, resulting in a wheel-like hybrid resonator (figure 1a-iii). Finally, an angled ion beam was used to efficiently create $V_B^-$ in the thin hBN flake. TRIM calculation (figure SI-2a) was performed to obtain the optimal condition for generation of defects within 20 nm of hBN using a nitrogen beam at an angle of 60° and fluences of $1\times10^{14}$ cm$^{-2}$.

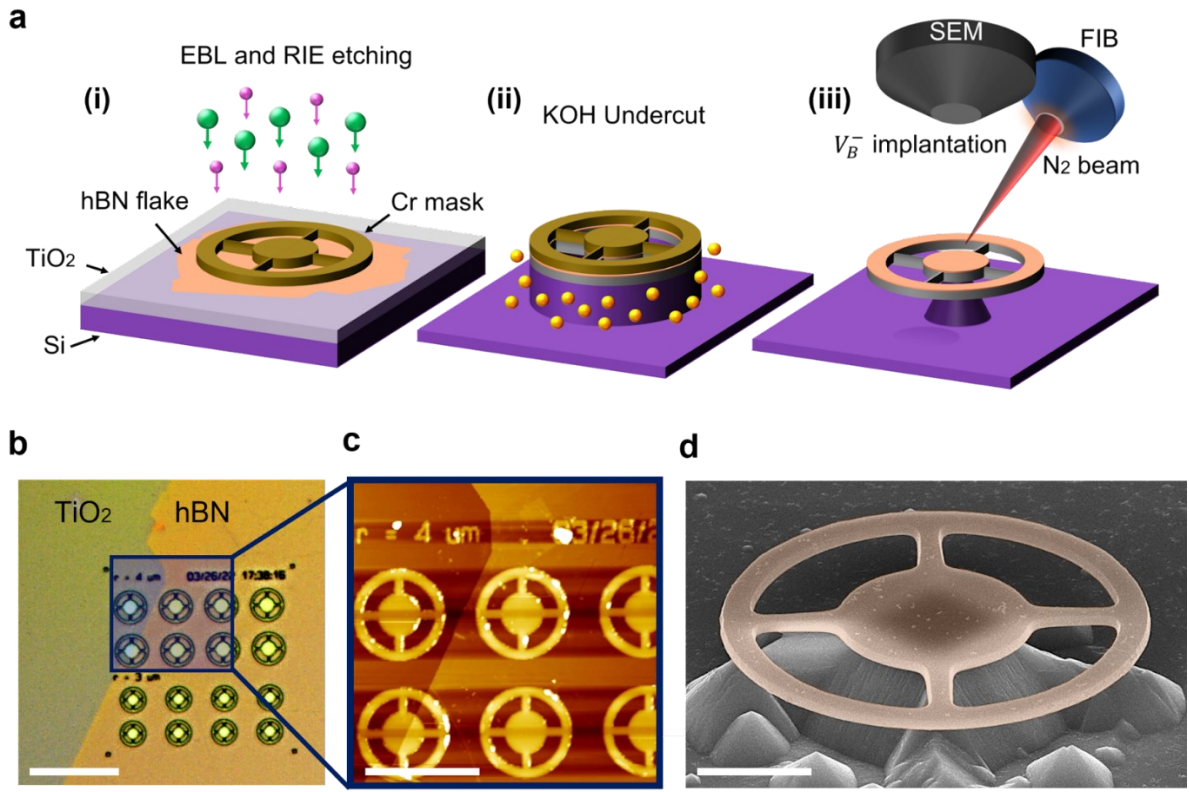

**Figure 1.** *Fabrication of hBN/TiO$_2$ hybrid resonator. (a) Schematic representation of the fabrication process: (i) Lithography step to pattern a Cr metal mask (Brown steering-wheels) on hBN/TiO$_2$ heterostructure and then transferring the mask into the heterostructure using ICP/RIE. Green and pink balls represent SF$_6$, and Hydrogen species used in the latter step, respectively. (ii) Undercut of hBN/TiO$_2$ hybrid resonator using 10% KOH aqueous solution. (iii) Generation of spin defects in hBN using angled Nitrogen focused ion beam (FIB). (b) Optical image of aligned EBL pattern on the hBN/TiO$_2$ heterostructure. Scale bar: 25 μm. (c) AFM image of the patterned area. scale bar: 10 μm. (d) 52° Tilted SEM image of the final hBN/TiO$_2$ hybrid resonator. The false color shows the microring. scale bar: 2 μm.*

Figure 1b shows an optical microscope image of the patterned sample. The spatially precise lithography on the flake is directly visible by optical contrast showing boundaries between hBN flake and the TiO$_2$ substrate. The Energy-dispersive X-ray spectroscopy (EDS) analysis after fabrication of the device also indicates the presence of hBN flake on top of the post-fabricated TiO$_2$ resonator (figure SI-3). An Atomic Force Microscopy (AFM) image was scanned over the patterned area to confirm the thickness of patterned hBN as shown in figure 1c and figure SI-2b. A detailed structure of the final hybrid resonator is shown in a high-resolution SEM image in figure 1d. Further SEM images of the resonators are presented in the supporting information indicating the quality of the resonator after fabrication (Figure SI-4).

Next, we characterized the coupling of $V_B^-$ to the TiO$_2$ resonator using a confocal photoluminescence setup (Method section). The photoluminescence spectrum collected from the center of the ring (denoted as "off-ring") is shown in figure 2a (blue spectra) and a relatively broad emission spectrum of $V_B^-$ spanning from 750 to 850 nm was observed. In contrast, when $V_B^-$ was excited on the ring (denoted as "on-ring", red spectra), superimposed peaks resulting from the coupling of $V_B^-$ emission to WGM of the rings were observed. Since the WGMs of the resonator cover entire $V_B^-$ emission range with short FSR of ~7 nm, overall $V_B^-$ signals were significantly enhanced compared to those of from uncoupled region. The coupled signals were collected from the scattering point (the bright spot indicated by red circle in the inset) where the photons partially outcoupled into the objective. A pristine TiO$_2$ ring resonator without hBN was also measured as a reference (figure SI-5). Normalized Electric (TE) and Magnetic (TM) field intensity was calculated using FDTD simulation as shown in figure 2b and figure SI-6. Electric field profile of the WGMs concentrated at the middle of the device whilst the magnetic field maxima confined at the top and bottom surface of the resonator as shown in the cross-sectional views. The analysis of the individual peak from high resolution spectra is shown in figure 2c. A quality factor ($Q = \lambda/\Delta\lambda$) of 2137 was obtained after fitting the peak with Lorentzian function. Characterization of more devices are shown in figure SI-7 and a mean quality factor of 1800 was achieved in this study.

To further investigate the enhancement of the coupled signals, PL intensity was measured for both on-ring and off-ring emissions as a function of excitation power shown in figure 2d. The intensities for both cases, were integrated over the whole range of $V_B^-$ emission, and the resulting data were fitted with the equation $I = (I_{sat}P)/(P + P_{sat})$, where $I_{sat}$ is the saturation intensity, and $P_{sat}$ is the saturation power. We determined saturation intensity of 0.405 MHz and 0.152 MHz for the on-ring and off-ring, respectively, and we observed the enhancement factor of about 3 and 7 in the saturated and unsaturated regimes.

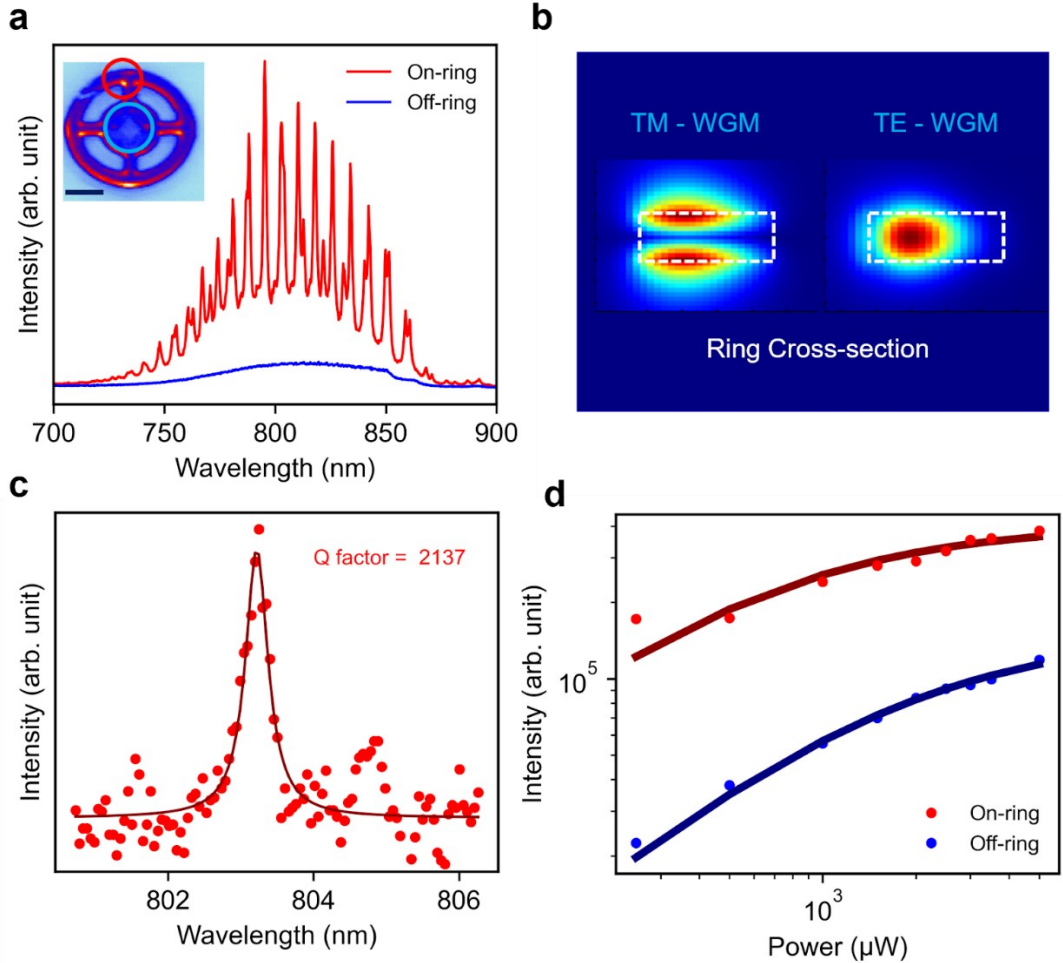

***Figure 2.*** *Optical characterization of the microring resonator. (a) PL spectrum collected on (red) and off (blue) the ring. Inset – confocal scanning image of the corresponding hybrid resonator. Red and blue circles indicate collection spots corresponding to On and Off the ring, respectively. Scale bar: 2 μm. (b) Cross-section view of the FDTD simulation for the fundamental TE and TM modes, showing magnetic and electric field distribution of the fundamental modes inside the resonator. The dash box outlines the $TiO_2$ cross-section view. (c) High-resolution spectra of the WGMs fitted by Lorentzian function yielding a quality factor of ~2100 for the fabricated ring resonator (d) Integrated PL intensity of the signals collected on (red) and off (blue) the ring at increasing excitation power.*

Ultimately, to further characterize the hybrid devices, we employed ODMR spectroscopy to map spin transitions of the coupled $V_B^-$ in local and non-local excitation configurations. The spin defect in hBN and its simplified electronic level structure is schematically depicted in figure 3a. The defect shows a triplet ground state ($A_1^3$) with as zero field splitting of 3.48 GHz between the two spin states. We evaluate spin transitions from $V_B^-$ that coupled to WGM in the local and non-local configurations. In the local configuration, the optical collection and excitation spots were overlapped, while at the non-local configuration, the excitation of the microscope was offset 180º on the ring away from the collection spot. The results and the schematic configuration of both scenarios are shown in figure 3b. The spectra recorded from non-local configuration is shown in figure 3b (purple) with similar features as the local excitation (green) indicating the efficient coupling of $V_B^-$ into the TiO$_2$ ring resonator. Figure 3c shows the corresponding ODMR signals for both excitation schemes. By sweeping a microwave from 3.2 to 3.8 GHz, a reduction in the photoluminescence of the $V_B^-$ was detected at 3.4 and 3.5 GHz. These contrasts are the result of transitions from m$_s$ = 0 to m$_s$ = ±1 in the ground state which are driven by the microwave field (figure 3a). This verifies that $V_B^-$ emits photons carrying spin information into the resonator and effectively coupled to the WGMs. Both excitation scheme showed similar ODMR contrast with slight differences stemming from lower collection efficiency through non-local excitation.

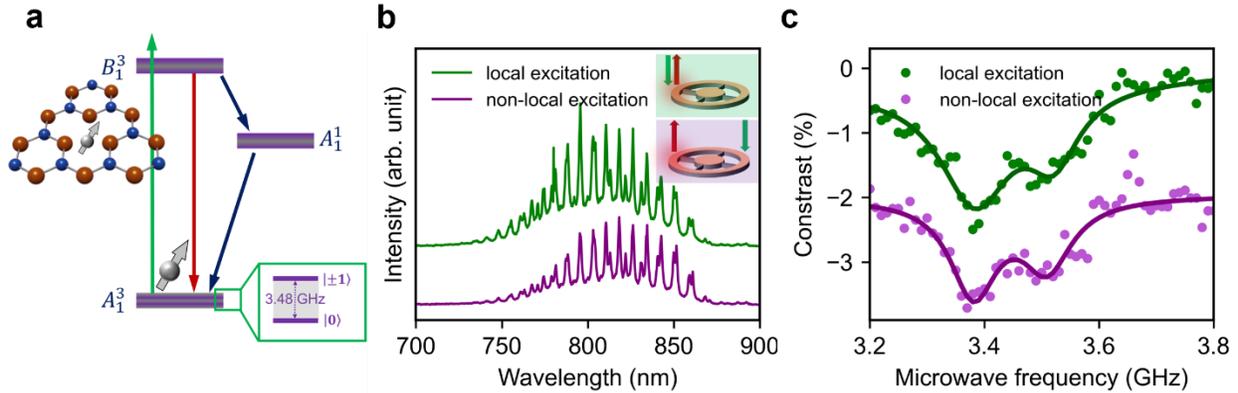

**Figure 3.** *ODMR measurement of coupled $V_B^-$. (a) Schematic of the $V_B^-$ defect and an energy diagram showing the main optical pumping cycle and the alternate cycle though the metastable state. (b) PL spectrum from coupled $V_B^-$ in local and non-local configurations. The inset schematically shows these configurations. (c) ODMR measurement from coupled $V_B^-$ in local and non-local configurations.*

In summary, we have developed a method for fabrication of a hBN/ TiO$_2$ resonators with high Quality factors exceeding ~ 2000. Coupling of $V_B^-$ to the WGM results in the increase of photoluminescence of about 7 and 3 in the unsaturated and saturated regimes, respectively. Finally, we demonstrated the coupling of $V_B^-$ via two excitation configurations. Importantly, in a non-local configuration, the PL intensity and the ODMR signal collected from guided photons, indicating efficient coupling of $V_B^-$ to TiO$_2$ ring resonators. Our results are promising for scaling up integration of layered materials with nanophotonic resonators.

**Acknowledgments**

The authors acknowledge financial support from the Australian Research Council (CE200100010) for financial support, the Asian Office of Aerospace Research and Development (FA2386-20-1-4014) and the Office of Naval Research Global (N62909-22-1-2028). The authors thank the ANFF (UTS node) for use of the nanofabrication facilities.